\documentclass[final]{agujournal2019}
\usepackage{url} 
\usepackage{lineno}
\usepackage[inline]{trackchanges} 
\usepackage{soul}
\usepackage{amssymb}
\usepackage{amsmath}
\definecolor{dred}{rgb}{0.7,0,0}
\definecolor{dgreen}{rgb}{0,0.5,0}
\definecolor{dblue}{rgb}{0,0,0.7}

\draftfalse

\journalname{Journal of Advances in Modeling Earth Systems (JAMES)}

\begin{document}

%
%


\title{Scale-awareness in an eddy energy constrained mesoscale eddy parameterization}

%
%




\authors{J. Mak\affil{1,2,3}, J. R. Maddison\affil{4}, D. P. Marshall\affil{5}, X. Ruan\affil{1}, Y. Wang\affil{1,2} and L. Yeow\affil{6}}


\affiliation{1}{Department of Ocean Science, Hong Kong University of Science and Technology}
\affiliation{2}{Center for Ocean Research in Hong Kong and Macau, Hong Kong University of Science and Technology}
\affiliation{3}{National Oceanography Centre, Southampton}
\affiliation{4}{School of Mathematics and Maxwell Institute for Mathematical
Sciences, The University of Edinburgh}
\affiliation{5}{Department of Physics, University of Oxford}
\affiliation{6}{Department of Physics, The Chinese University of Hong Kong}




\correspondingauthor{Julian Mak}{julian.c.l.mak@googlemail.com}




\begin{keypoints}
\item Evidence is presented for scale-awareness of the GEOMETRIC eddy parameterization with a `splitting' approach
\item Scale-awareness means constancy of total eddy energy, while the explicit and parameterized components vary with resolution
\item In eddy permitting calculations, such eddy parameterizations should be applied to the large-scale background state, obtained via `splitting'

\end{keypoints}

%
%

%
%


\begin{abstract}
There is an increasing interest in mesoscale eddy parameterizations that are
\emph{scale-aware}, normally interpreted to mean that a parameterization does
not require parameter recalibration as the model resolution changes. Here we
examine whether Gent--McWilliams (GM) based version of GEOMETRIC, a mesoscale
eddy parameterization that is constrained by a parameterized eddy energy budget,
is scale-aware in its energetics. It is generally known that GM-based schemes
severely damp out explicit eddies, so the parameterized component would be
expected to dominate across resolutions, and we might expect a negative answer
to the question of energetic scale-awareness. A consideration of why GM-based
schemes damp out explicit eddies leads a suggestion for what we term a
\emph{splitting} procedure: a definition of a `large-scale' field is sought, and
the eddy-induced velocity from the GM-scheme is computed from and acts only on
the large-scale field, allowing explicit and parameterized components to
co-exist. Within the context of an idealized re-entrant channel model of the
Southern Ocean, evidence is provided that the GM-based version of GEOMETRIC is
scale-aware in the energetics as long as we employ a splitting procedure. The
splitting procedure also leads to an improved representation of mean states
without detrimental effects on the explicit eddy motions.
\end{abstract}

\section*{Plain Language Summary}
With increasing computational power, ocean models are starting to explicitly
resolve eddy motions with characteristic length-scales of 10 to 100 km. With the
increased model resolution, there is an increasing call for \emph{scale-aware
parameterizations}, i.e. simplified models that are supposed to represent the
missing eddy feedbacks onto the modeled state that are either self-tuning, or do
not require re-tuning of parameters, without damping explicit eddies resolved by
the model. The question we ask is whether an eddy parameterization known as
GEOMETRIC is scale-aware. The expected answer might be ``no'': such schemes are
known to heavily flatten density variations and damp out explicitly resolved
eddies. We instead propose a \emph{splitting} scheme that avoids damping of
explicitly resolved eddies. The GEOMETRIC scheme, with the use of splitting,
seems to be scale-aware in the sense that there is approximate constancy of
total eddy energy, where the explicit and parameterized components vary with
resolution, and additionally lead to other desirable features in the model
considered.

%
%

%


%
%
%
%





\section{Introduction}\label{sec:intro}

The ocean is a central component of the Earth system, and changes in the
overturning circulation have important consequences for the global energy and
biogeochemical cycles \cite<e.g.,>{ZhangVallis13, Adkins13, Ferrari-et-al14,
Burke-et-al15, Bopp-et-al17, Takano-et-al18, GalbraithdeLavergne19, Li-et-al20}.
In particular, it is known that the representation of geostrophic mesoscale eddy
processes in numerical models, whether explicitly permitted by the numerical
model or through a parameterization, can have a significant influence on the
ocean climate sensitivity \cite<e.g.,>{FoxKemper-et-al19, Hewitt-et-al20}.

Models employing parameterizations often require ``tuning'', or parameter
calibration, in that the paramaeters associated with the parameterizations are
adjusted so that the model reduces known biases of the model, such as sea
surface temperature patterns, overall Southern Ocean volume transport, or mixed
layer depths, relative to a high resolution model truth and/or observational
data. On the other hand, with increasing computational power, there is an
increasing possibility for numerical ocean models to increase in spatial
resolution. With an increased horizontal resolution the geostrophic mesoscale
motion starts becoming explicitly represented, which is known to offer some
benefits over parameterizations, such as improvements to global/regional biases
in the relevant ocean metrics, such as those mentioned above
\cite<e.g.,>{Hewitt-et-al17, Hewitt-et-al20}. While there is an increasing push
for the so called $k$-scale (kilometer scale) models that aim to explicitly
resolve the eddy dynamics and dispense with parameterizations altogether
\cite{Slingo-et-al22, Hewitt-et-al22}, such models are computationally
prohibitive for climate applications, which tend to require long-time
integrations. A more realistic and achievable goal in the near future is for
ocean models used for climate projections to be eddy permitting, at roughly
$1/4^\circ$ horizontal resolution, where the modeled eddy-mean feedback is known
to be insufficient, and some form of parameterization is still required for the
missing eddy feedbacks. It is generally beneficial to have a traceable hierarchy
of numerical models differing mostly in the resolution
\cite<e.g.,>{Storkey-et-al18}, to tackle the appropriate questions in a
computationally tractable manner depending on the time or spatial scales of
interest.

With such a hierachy of models, there is an increasing interest in
\emph{scale-aware} parameterizations, in that one set of parameter choice is
applicable across model resolutions. Examples of these include parameterizations
based on Large Eddy Simulations designed for rotating stratified turbulence,
with a resulting eddy viscosity/diffusion that is grid-scale dependent
\cite<e.g.,>{Smagorinsky93, Bachman-et-al17b}. On the other hand, more
traditional parameterizations, such as the Gent--McWilliams (GM) scheme
\cite{GentMcWilliams90, Gent-et-al95}, isoneutral diffusion \cite<e.g.,>{Redi82,
Griffies-et-al98}, enhanced vertical momentum diffusion
\cite<e.g.,>{GreatbatchLamb90}, and backscatter \cite<e.g.,>{Zanna-et-al17,
Bachman19, Jansen-et-al19, Juricke-et-al20}, are based on Reynolds averaging
procedures. The first three cases and variants thereof normally rely on a
specification of a diffusivity or a transfer coefficient and are generally not
scale-aware by construction, although there are proposals to make these schemes
scale-aware, such as via a resolution function that scales the diffusivity
according to the modeled state \cite<e.g.,>{Hallberg13}. Backscatter schemes,
where a portion of the energy dissipated at the grid-scale or other means is
re-injected back into the resolved scales, is scale-aware in the sense that the
scheme itself dynamically adjusts depending on the model length-scales (be it
the grid-scale, effective resolution based for example on the Rossby deformation
radius, or otherwise).

\begin{figure}
  \includegraphics[width=\textwidth]{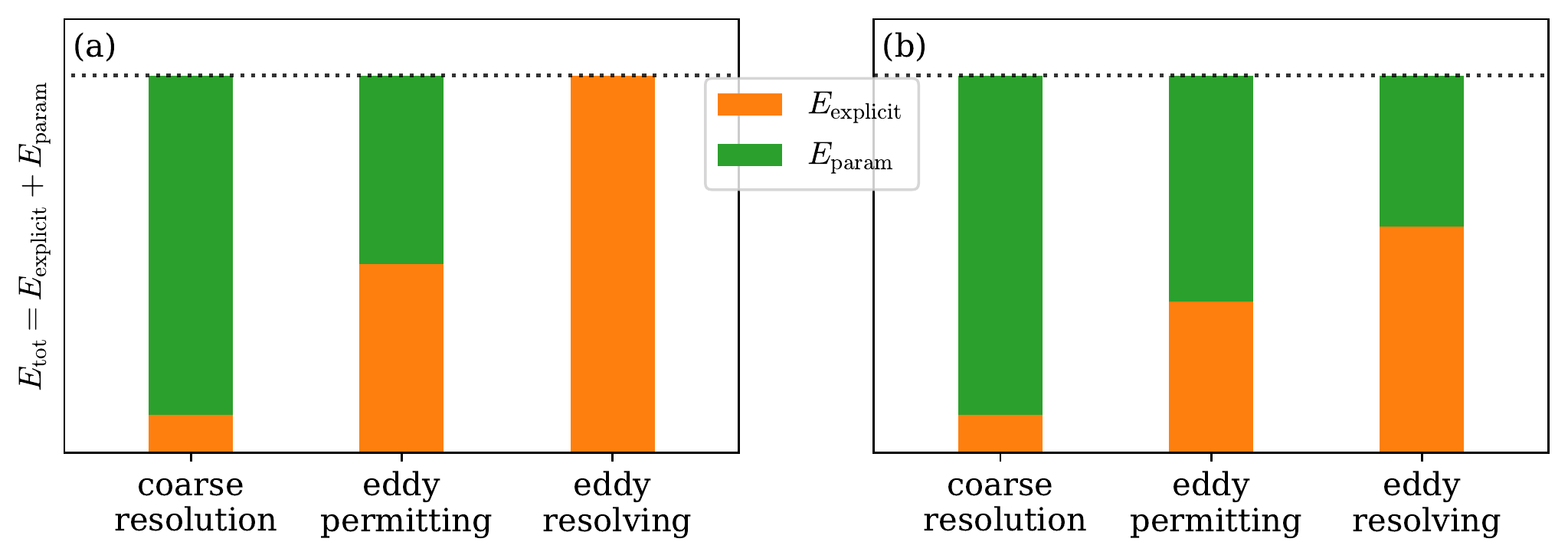}
  \caption{Possible scenarios for scale-awareness in the total (explicit and
  parameterized, orange and green respectively) eddy energy. ($a$) An ideal
  situation where, only tuning for the parameterized eddy energy component in
  the coarse resolution model, the parameterized eddy energy component decreases
  as resolution increases, eventually switching off completely, such that the
  total eddy energy remains constant. ($b$) A more likely but still desirable
  scenario, where the parameterized eddy energy component does not completely
  vanish, but the total eddy energy remains constant.}
  \label{fig:ene_decomp_schematic}
\end{figure}

The principal focus of this work is on GM-based parameterizations, valid for
geostrophic mesoscale motions for which the Rossby number is small. With the
increased prevalence of eddy energy constrained GM-based parameterizations
\cite<e.g.,>{EdenGreatbatch08, MarshallAdcroft10, Marshall-et-al12,
Jansen-et-al19}, an aspect that we consider in this work is scale-awareness in
the total eddy energy. We suppose that there is a fixed amount of total eddy
energy available, but this can be represented in explicit or parameterized
forms. The question here is whether an energetically constrained eddy
parameterization is scale-aware in the eddy energetics, in the sense that total
(explicit and parameterized) eddy energy is conserved as model resolution is
varied without re-tuning of parameters, while the partition into explicit and
parameterized components may vary. An ideal scenario we envisage would be where
the parameterized eddy energy decreases while the explicit component increases
as resolution increases, such that the sum remains constant, as illustrated in
Fig.~\ref{fig:ene_decomp_schematic}($a$). The eddy resolving model has all the
eddy energy in the explicit component, and explicit eddies are responsible for
all the eddy-mean interaction. The coarse resolution models are tuned so that
the total eddy energy is roughly the same as the eddy resolving model, but most
of the eddy energy contribution is in the parameterized component. Without
additional tuning, as we go into the eddy permitting regime, some of that work
by parameterized eddies are taken up by the explicit eddies, but such that the
total eddy energy remains constant. This could be possible with energetically
constrained parameterizations, since the resolved modeled state information is
often utilized in the parameterized eddy energy budgets, but by no means
guaranteed. An alternative and more plausible (but still desirable) scenario
might be that illustrated in Fig.~\ref{fig:ene_decomp_schematic}($b$), where the
parameterized component reduces in magnitude but does not completely vanish as
we move from the coarse through eddy permitting to eddy resolving resolutions.
Without further tuning, there may still be a non-negligible amount of eddy
energy in the parameterized component, but this might be regarded as scale-aware
since the total eddy energy level is still roughly constant.

\begin{figure}
  \includegraphics[width=\textwidth]{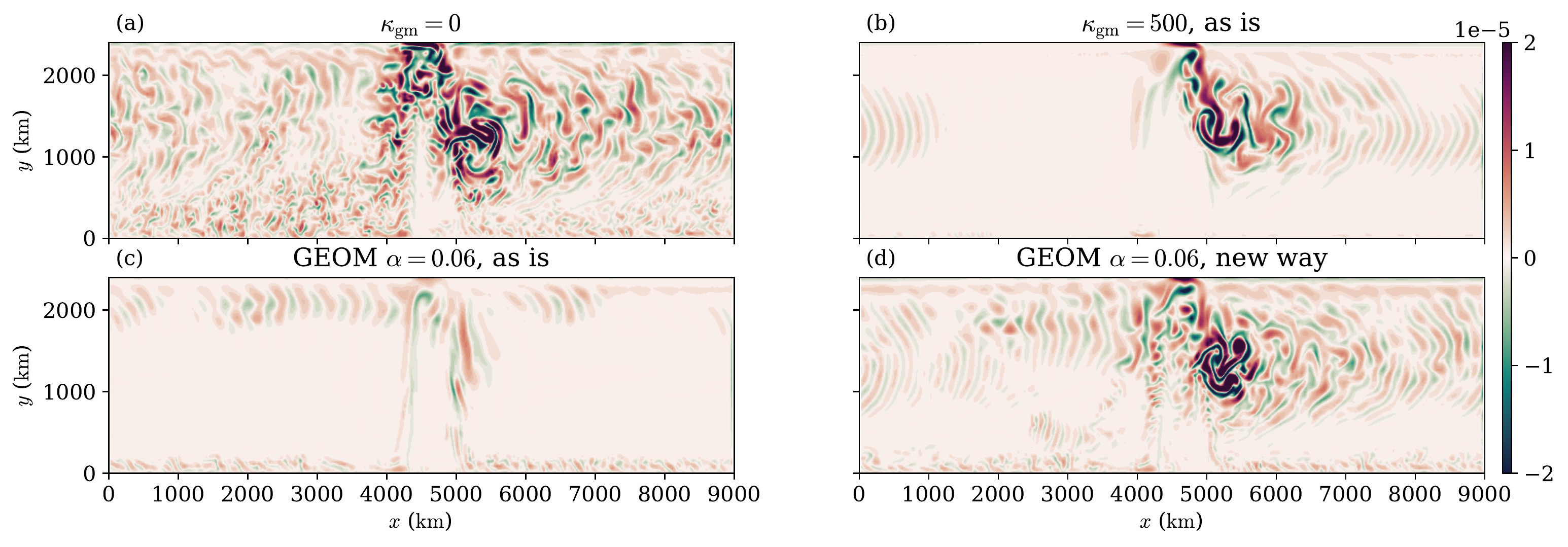}
  \caption{Snapshot of the surface relative vorticity (in units of
  $\mathrm{s}^{-1}$) for the $25\ \mathrm{km}$ horizontal resolution re-entrant
  channel calculation, at the end of model year 810 (model year 501 being the
  start of the perturbation experiments). ($a$) $\kappa_{\rm gm} = 0\
  \mathrm{m}^2\ \mathrm{s}^{-1}$. ($b$) $\kappa_{\rm gm} = 500\ \mathrm{m}^2\
  \mathrm{s}^{-1}$. ($c$) GEOMETRIC prescription of $\kappa_{\rm gm}$ applied as
  standard. ($d$) GEOMETRIC prescription of $\kappa_{\rm gm}$ with approach
  detailed in this article. The domain average of $\kappa_{\rm gm}$ for panels
  ($c,d$) are diagnosed to be around $2500$ and $750\ \mathrm{m}^2\
  \mathrm{s}^{-1}$ respectively.}
  \label{fig:xi_snap}
\end{figure}

For this work we focus on one of the existing eddy energy constrained mesoscale
eddy parameterizations, the GM-based version of GEOMETRIC
\cite<e.g.,>{Marshall-et-al12, Mak-et-al18}, which has been implemented and
tested in various numerical ocean general circulation models in a variety of
configurations \cite<e.g.,>{Mak-et-al18, Mak-et-al22, Ruan-et-al23}. The
principal question we address in the present article is whether GEOMETRIC,
without the use of resolution functions or backscatter, is itself scale-aware in
terms of the eddy energetics. A reasonable first guess at the answer would be
``no'': the GM-based schemes are intended for calculations at non-eddying
resolutions, and are known to severely damp any mesoscale eddies that are
explicitly represented in eddy permitting models. An example of this excessive
damping is shown in Fig.~\ref{fig:xi_snap}($b,c$), where the use of GM-based
schemes as standard in a 25 km horizontal resolution model (to be introduced in
\S\ref{sec:method}) severely damps the instantaneous fluctuations, as
represented by the surface relative vorticity, relative to
Fig.~\ref{fig:xi_snap}($a$) where no GM-based scheme is active. The use of the
GM-based schemes as-is puts us firmly in the regime where the parameterized
component is almost entirely responsible for the eddy-mean interaction and
occupies a large percentage of the total eddy energy, and in this instance we
would not expect scale-awareness.

In \S\ref{sec:method} we review the reasons why GM-based schemes damp explicit
eddies, suggesting what we term here a \emph{splitting} procedure that would
enable GM-based schemes (not necessarily just GEOMETRIC) to avoid damping
explicit eddy activity; see Fig.~\ref{fig:xi_snap}($d$) for an example of
GEOMETRIC employing the splitting procedure. A splitting procedure puts us in
the regime where parameterized and explicit eddies may in principle co-exist
without intruding on one another. Being constrained by an overall eddy energy
budget that depends on the resolved mean-state, GEOMETRIC could be scale-aware
in the eddy energetics. The presence or absence of explicit eddies would lead to
variations in the resolved mean-state that is expected to impact the
parameterized eddy energy. The parameterized eddy feedback onto the resolved
mean-state is dependent on the parameterized eddy energy, and the parameterized
eddy feedback might adjust according to the level of explicit eddy feedback. The
exact implementation of the splitting procedure and the numerical experiments
for testing whether we have energetic scale-awareness in GEOMETRIC are detailed
in \S\ref{sec:model}. \S\ref{sec:results} provides evidence in support of
GEOMETRIC having scale-aware energetics, and further examines other benefits
offered by the splitting procedure, focusing on the model in the eddy permitting
regime with a horizontal resolution of 25 km (roughly $1/4^\circ$). We close the
article in \S\ref{sec:conc} where we discuss the results, provide outlooks, as
well as the practical and modeling implication for the splitting procedure
detailed in this work, such as the use of a resolution function and backscatter
parameterizations.


\section{The GM scheme and the splitting approach}\label{sec:method}


\subsection{GM scheme currently as applied}

The GM scheme, while resembling a horizontal buoyancy diffusion \cite<in the
quasi-geostrophic limit, e.g.,>{Treguier-et-al97} or a layer thickness diffusion
\cite<e.g.,>{GentMcWilliams90}, is really an advection
\cite<e.g.,>{Gent-et-al95, Treguier-et-al97, Griffies98, Ferreira-et-al05}, and
introduces an eddy-induced velocity $\boldsymbol{u}^*$ as
\begin{linenomath*}
\begin{equation}\label{eq:eiv}
  \boldsymbol{u}^* = \nabla\times(\boldsymbol{e}_z \times \kappa_{\rm gm} \boldsymbol{s}), \qquad \boldsymbol{s} = \frac{\nabla_H \rho}{-\partial \rho / \partial z}.
\end{equation}
\end{linenomath*}
Here, $\boldsymbol{e}_z$ denotes the unit vector pointing in the vertical
direction, $\boldsymbol{s} = (s_x, s_y, 0)$ encodes the isopycnal slopes in the
horizontal directions, $\rho$ is the dynamically relevant density, $\nabla_H$
the horizontal gradient operator, and $\kappa_{\rm gm}$ is the GM or the
eddy-induced velocity coefficient. The GM scheme only adds the eddy-induced
velocity to the tracer equations: assuming that the only thermodynamic variable
is the temperature $\Theta$, the active tracer equation is modified to
\begin{linenomath*}
\begin{equation}\label{eq:traadv}
  \frac{\partial \Theta}{\partial t} + (\boldsymbol{u} + \boldsymbol{u}^*) \cdot \nabla \Theta = \ldots,
\end{equation}
\end{linenomath*}
where $\boldsymbol{u}$ is the resolved velocity, and the right-hand-side
includes the relevant forcing, diffusion and/or dissipation terms. The
eddy-induced velocity as represented by the GM scheme acts to flatten
isopycnals, originally intended to mimic the action of baroclinic instability.
In practice it ends up leading to a parameterized interfacial form stress
regardless of the generating mechanism, since meridional advection of
buoyancy/density is equivalent to a vertical transfer of horizontal momentum in
the quasi-geostrophic limit \cite<e.g.,>{GreatbatchLamb90, McWilliamsGent94,
Marshall-et-al12}.

The GM scheme was originally designed for models with no explicit eddies, where
the isopycnal slope $\boldsymbol{s}$ is a fundamentally large-scale field with
no small-scale fluctuations in both the velocity and density (the two variables
being related via the thermal wind shear relation). Having GM-based schemes
switched on when explicit eddies are permitted severely damps the explicit
eddies. The reason for this is essentially given in Eq.~(\ref{eq:eiv}): with
explicit eddies in regimes where thermal wind balance holds, the resulting
isopycnal slope $\boldsymbol{s}$ has small-scale fluctuations, so that the
resulting eddy-induced velocity $\boldsymbol{u}^*$ is potentially large
magnitude at small-scales, via a curl of $\boldsymbol{s}$. The resulting
$\boldsymbol{u}^*$ acts to rapidly damp the smaller-scale explicit fluctuations
that are permitted. If the magnitude of $\kappa_{\rm gm}$ is large, the
resulting model calculation strongly resembles a coarse resolution model with no
explicit eddies (cf. Fig.~\ref{fig:xi_snap}$c$, and sample calculations not
shown with spatially constant but large $\kappa_{\rm gm}$). On the other hand,
while the calculation with no GM-based scheme switched on looks better and
possesses stronger explicit fluctuations, the mean state turns out to be
slightly problematic, attributed to the explicit eddy feedback onto the
mean-state being too weak (e.g., the stratification and circumpolar transport
associated with the Fig.~\ref{fig:xi_snap}$a$ calculation will be seen to be too
deep and too strong respectively). It would appear that some form of
parameterization is required to supplement the missing eddy-mean interaction.

Several approaches have been proposed to combat the excessive damping and to
increase the eddy impact on the mean state, and are sometimes used in
combination. One is to consider an anisotropic version of the GM
parameterization, where the anisotropy refers to the along and across stream
direction \cite<e.g.,>{SmithGent04}, although this is not commonly implemented.
Another is to control the value of $\kappa_{\rm gm}$ based on a grid spacing
and/or the Rossby deformation radius of the resolved state
\cite<e.g.,>{Hallberg13}, and the GM scheme is thus only functioning where the
state is regarded as `under-resolved'. Yet another is via a momentum-based
backscatter approach \cite<e.g.,>{Zanna-et-al17, Bachman19, Jansen-et-al19,
Juricke-et-al20}, which aims to model the pathway of energy flowing towards
large-scales associated with rotationally constrained turbulence
\cite<e.g.,>{Charney71, Rhines75, Salmon80, VallisMaltrud93, SrinivasanYoung12,
WatermanJayne12}, and has the benefit of energizing the explicit flow. One
possible critique with the use of a resolution function is that controlling
$\kappa_{\rm gm}$ affects the magnitude but not the small-scale nature of
$\boldsymbol{u}^*$, and the GM scheme is still applied outside of its intended
domain of validity. A possible criticism of the damping first then backscatter
approach is that it risks compensating overly strong dissipation with overly
strong backscatter, leading to two balancing or competing unphysical mechanisms.
For example, the strong small-scale dissipation of a direct application of GM at
eddy permitting resolution does not correspond to large-scale baroclinic
instability, but instead to the parameterization being used outside of its
intended regime of validity. While physical eddy backscatter is observed and is
a target for parameterization, it should not be confused with numerical
backscatter acting to counter excessive dissipation.


\subsection{A field splitting approach}

\begin{figure}
  \begin{center}\includegraphics[width=0.9\textwidth]{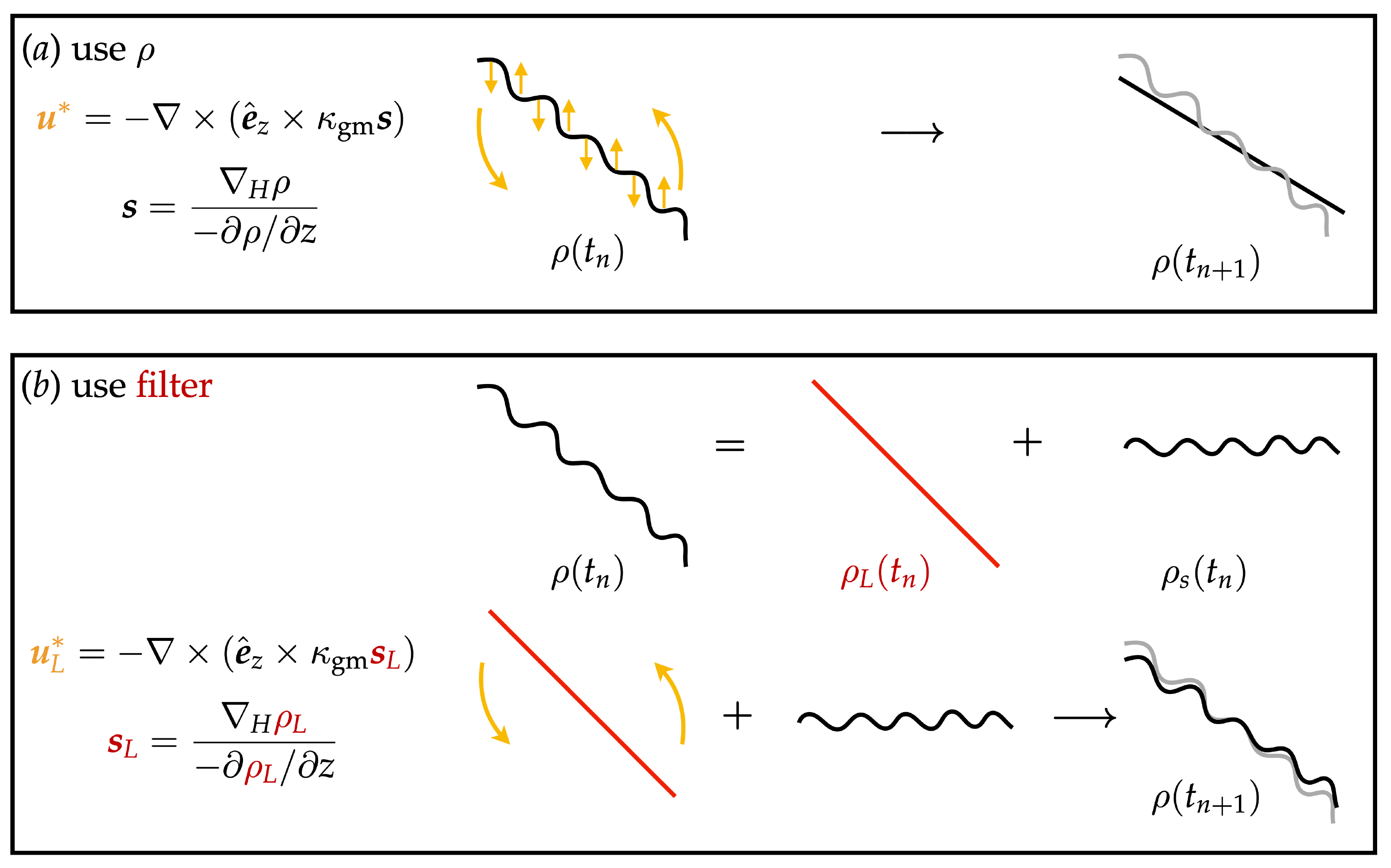}\end{center}
  \caption{Schematic comparing existing and present approach. (Top) Existing
  approach, where the isopyncal slopes $\boldsymbol{s}$ and so
  $\boldsymbol{u}^*$ is a small-scale field. (Bottom) Present proposal, where we
  ask for a definition of a large-scale density field, so that the associated
  $\boldsymbol{s}_L$ and $\boldsymbol{u}^*_L$ are both large-scale, and
  $\boldsymbol{u}^*_L$ would only act on the large-scale field. Black denotes
  the isopycnals associated with the full/small-scale state, orange the eddy
  induced velocity at the relevant scales, and red the isopycnals associated
  with large-scales.}
  \label{fig:schematic}
\end{figure}

Given the previous discussion alluding to large/small-scale fields, the proposal
here is to consider an approach motivated by the schematic given in
Fig.~\ref{fig:schematic}. Given an appropriate definition of a large-scale
density field $\rho_L$, where $\rho = \rho_L + \rho_S$, we compute the
associated large-scale isopycnal slope $\boldsymbol{s}_L = \nabla_H \rho_L /
(-\partial \rho_L / \partial z)$. The associated eddy-induced velocity is
computed using $\boldsymbol{s}_L$, i.e.
\begin{linenomath*}
\begin{equation}\label{eq:eiv_L}
  \boldsymbol{u}^*_L = \nabla\times \left(\boldsymbol{e}_z \times \kappa_{\rm gm} \boldsymbol{s}_L\right),
\end{equation}
\end{linenomath*}
and we might apply this $\boldsymbol{u}^*_L$ only to the large-scale active
tracer equations.

This proposal might be expected to address the critique that GM is being used
beyond its regime of validity and the excessive damping that is observed when
using GM in eddy permitting regimes. Since $\boldsymbol{s}_L$ is a large-scale
field by construction, the resulting $\boldsymbol{u}^*_L$ will also be
large-scale, at least much larger-scale than the scales associated with the
explicit eddies using a suitable definition. As a corollary, since
$\boldsymbol{u}^*_L$ is a curl of a large-scale rather than small-scale field,
we would expect $|\boldsymbol{u}^*_L| \ll |\boldsymbol{u}^*|$. The present
proposal thus controls \emph{both} the spatial distribution and the magnitude of
the eddy-induced velocity, and is perhaps closer in spirit to the original
intention of the GM scheme \cite{GentMcWilliams90, Gent-et-al95}, that is, a
parameterization on a background that has no eddies, here defined by a
large-scale field that has the explicit eddy field filtered out in some way.

The present proposal supplements the explicit interfacial form stress with some
parameterized form stress computed only using the large-scale field, but in a
way that aims to avoid cancellations through competing effects. By comparison,
controlling $\kappa_{\rm gm}$ aims to avoid excessive dissipation completely,
while GM-based schemes with backscatter would dissipate first and, accepting
that there is cancellation, add back some of the missing parts. That said, the
present proposal is not mutually exclusive of additional controls on
$\kappa_{\rm gm}$, nor on the use of backscatter. We would argue that, with the
present splitting approach, no control on $\kappa_{\rm gm}$ is strictly
necessary, since this is already taken care of by the use of
$\boldsymbol{u}^*_L$. Already damping fewer eddies, such a scheme would imply
that, when used in conjunction with backscatter, the degree of backscatter might
not need to be that substantial.

The splitting approach here does not require substantial modification of the
relevant GM-based schemes themselves, since only the data used in computing the
relevant quantities are modified. There are of course various theoretical and
practical aspects to consider, such as (1) the choice of filter, (2) the
relevant large-scale tracer equations to implement, (3) issues with nonlinear
equation of state, (4) computational cost issues, and others. These are further
discussed in \S\ref{sec:model} where we detail the precise implementation
choices considered in this work, and in \S\ref{sec:conc} where we evaluate of
our choices and provide possible alternative.


\subsection{Parameterized eddy energetics}\label{sec:energetics}

The splitting approach detailed above makes no specific choice of GM-based
scheme, and to address the question of whether there is eddy energy
scale-awareness, in that the total (explicit and parameterized) eddy energy
remains somewhat constant with changing model resolution, we focus our attention
on the GM-based version of the GEOMETRIC scheme \cite{Marshall-et-al12,
Mak-et-al18, Mak-et-al22}. The GM-version of GEOMETRIC arises from a bound via
analyzing the Eliassen--Palm flux tensor \cite{Marshall-et-al12,
MaddisonMarshall13} in the quasi-geostrophic regime, and results in a
$\kappa_{\rm gm}$ with a linear dependence on the total eddy energy $E$. The
implementation of \citeA{Mak-et-al22} in a global configuration ocean global
circulation model takes
\begin{linenomath*}
\begin{equation}\label{eq:geom}
  \kappa_{\rm gm} = \alpha \frac{\int E\; \mathrm{d}z}{\int (M^2 / N)\; \mathrm{d}z},
\end{equation}
\end{linenomath*}
where $E$ is the total (potential and kinetic) parameterized eddy energy,
prognostically determined by the parameterized eddy energy budget
\begin{linenomath*}
\begin{equation}\label{eq:ene-eq}
  \frac{\mathrm{d}\hat{E}}{\mathrm{d}t}
    + \underbrace{\nabla_H \cdot \left( \left(\widetilde{\boldsymbol{u}}^z - |c|\, \boldsymbol{e}_x\right) \hat{E} \right)}_\textnormal{advection}
   = \underbrace{\int \kappa_{\rm gm} \frac{M^4}{N^2}\; \mathrm{d}z}_\textnormal{source}
    - \underbrace{\lambda (\hat{E} - \hat{E_0})}_\textnormal{dissipation} 
    + \underbrace{\eta_E\nabla^2_H  \hat{E}}_\textnormal{diffusion}.
\end{equation}
\end{linenomath*}
Here, the depth-integrated total parameterized eddy energy $\hat{E} = \int E\;
\mathrm{d}z$ is advected by the depth-averaged flow
$\widetilde{\boldsymbol{u}}^z$, with westward propagation at the long Rossby
wave phase speed $|c|$ \cite<e.g.,>{Chelton-et-al11, KlockerMarshall14}. The
growth of parameterized eddy energy comes from the slumping of mean density
surfaces, where $M^2 = |(-g/\rho_0)\nabla_H\rho|$ and $N^2 =
-(g/\rho_0)\partial\rho/\partial z$ are the horizontal and vertical buoyancy
frequencies (so $M^2 / N^2 = s = |\boldsymbol{s}|$). The parameterized eddy
energy is diffused in the horizontal \cite{Grooms15, Ni-et-al20b, Ni-et-al20a}
with a diffusivity $\eta_E$. A linear dissipation of the parameterized eddy
energy at rate $\lambda$ is employed (with minimum parameterized eddy energy
level $E_0$), so $\lambda^{-1}$ is an time-scale, and is a bulk parameterization
of energy fluxes out of the mesoscales resulting from numerous dynamical
processes \cite<e.g.,>{Mak-et-al22b}.

One question here is on the nature of the quantities to be used in
Eq.~(\ref{eq:geom}) and Eq.~(\ref{eq:ene-eq}) if we consider employing a
splitting approach detailed above. To maintain a positive-definite source term,
we utilize
\begin{linenomath*}
\begin{equation}\label{eq:GEOMloc-e}
  \frac{\mathrm{d}\hat{E}}{\mathrm{d}t}
    + \underbrace{\nabla_H \cdot \left( \left(\widetilde{\boldsymbol{u}}^z - |c|\, \boldsymbol{e}_x\right) \hat{E} \right)}_\textnormal{advection}
   = \underbrace{\int \kappa_{\rm gm} \frac{M_L^4}{N_L^2}\; \mathrm{d}z}_\textnormal{source}
    - \underbrace{\lambda (\hat{E} - \hat{E_0})}_\textnormal{dissipation} 
    + \underbrace{\eta_E\nabla^2_H  \hat{E}}_\textnormal{diffusion},
\end{equation}
\end{linenomath*}
where the subscript $L$ denotes the quantities computed using the large-scale
density field $\rho_L$, and the form of the source term is supported by an
analysis analogous to that in Appendix A of \citeA{Mak-et-al17a}. While there is
no obvious restriction as such on the computation of $\kappa_{\rm gm}$ in
Eq.~(\ref{eq:geom}), it would be more consistent that we also use large-scale
information, i.e., in employing $\boldsymbol{u}^*_L$ as given in
Eq.~(\ref{eq:eiv_L}), we also compute
\begin{linenomath*}
\begin{equation}\label{eq:GEOM-gm}
  \kappa_{\rm gm} = \alpha\frac{\hat{E}}{\int M_L^2 / N_L\; \mathrm{d}z},
\end{equation}
\end{linenomath*}
so that the resulting $\kappa_{\rm gm}$ is a large-scale field.


\section{Implementation and model set up}\label{sec:model}

As a first implementation to test out the idea of scale-awareness in the eddy
energetics and the splitting approach, we make some simplifications to what was
proposed in \S\ref{sec:method} and Fig.~\ref{fig:schematic}. Specifically, we
consider (1) a filtering performed per horizontal level, and (2) the resulting
$\boldsymbol{u}^*_L$ is applied to the full tracer equation. The first is not an
unreasonable first approximation given we are dealing with systems with small
aspect ratios, and we normally expect the isopycnal slopes to be rather small.
While we would ideally like to define a large-scale based on some sort of
isopycnal-based averaging, a horizontal average at fixed depth is maybe a
reasonable first attempt, at least numerically. The second is more of a
theoretical issue, where we would in this instance like to derive an equation
only for the large-scale active tracer field, which is not an immediately
obvious endeavor because of the definition of the velocity; some ideas and
discussions are provided in \S\ref{sec:conc}. If however it is the case that the
use of a large-scale field already leads to a large-scale and small magnitude
$\boldsymbol{u}^*_L$, we might expect that the small-scales are advected in a
somewhat passive manner by $\boldsymbol{u}^*_L$, and it is possible that in
practice the expected reduction in damping may already allow the scale-awareness
in eddy energetics (if it exists) to emerge.


\subsection{Implementation in NEMO}

A natural way to filter fields might be to consider a diffusion-based filter.
Taking the tracer variable to be temperature $\Theta$ for concreteness, consider
the diffusion equation discretized in time given by
\begin{linenomath*}
\begin{equation}
  \Theta^{n+1} - \Theta^n = L^2 \nabla_H^2 \Theta,
\end{equation}
\end{linenomath*}
where $L^2 = \kappa \Delta \tilde{t}$ denotes a length-scale squared, $\kappa$
is a nominal diffusivity, and $\Delta \tilde{t}$ would be a nominal time-step
size. Here the superscripts on $\Theta$ denote the pseudo-time index, and the
idea here is to pseudo-time-step a full field into some large-scale field as
dictated by the diffusion operator. If we take the right hand side to be at
pseudo-time index $n$, then we are dealing with an explicit scheme that, while
easy to code up, is subject to strong constraints on the choice of $L$ through
stability conditions. One could consider repeated cycling via $\Theta^{n+1} = (1
+ L^2\nabla_H^2)^M\Theta^n$ for some $M \geq 1$, but in practice $M$ needs to be
quite large for finer resolution models because the choice of stable $L$ is
rather small, and we lose the interpretability of $L$ as a length-scale.

\begin{figure}
  \includegraphics[width=\textwidth]{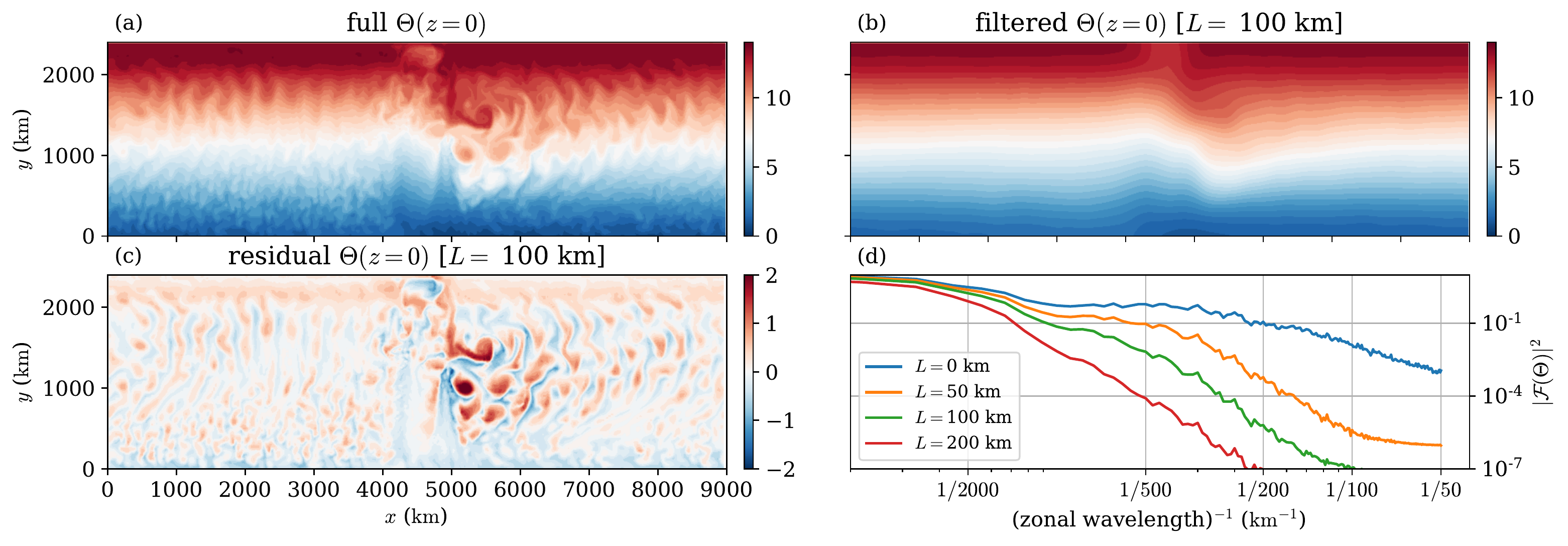}
  \caption{Demonstration of the filter action on a diagnosed snapshot of the
  temperature field from a $\Delta x = \Delta y = 25$ km calculation with
  $\kappa_{\rm gm} = 0$, with $M=2$. ($a$) The original field. ($b$) The
  filtered field with filter length $L=100$ km. ($c$) The small-scale field as a
  residual between the original and the filtered field (i.e., panel $a$ minus
  $b$). ($d$) Zonal power spectrum of the large-scale field at fixed latitude
  averaged over all latitudes (units of ${}^\circ\mathrm{C}^2$) for sample
  choices of $L$, demonstrating a drop off in the power spectrum to below
  $10^{-7}$ coinciding with choice of $L$.}
  \label{fig:pre_mortem_L100_R025}
\end{figure}

We consider instead
\begin{linenomath*}
\begin{equation}\label{eq:diff_imp}
  (1 - L^2\nabla_H^2)^2\Theta^{n+1} = \Theta^n,
\end{equation}
\end{linenomath*}
which bears resemblance to solving the diffusion equation by a backward Euler
scheme, except with the $2$ exponent on the operator $(1 - L^2\nabla_H^2)$.
Dealing with implicit schemes alleviates the constraint placed by stability
conditions and $L$ can in principle be chosen independent of the choice of model
resolution. The operator in this case is symmetric positive-definite, so there
are a variety of solvers available \cite<e.g., conjugate
gradient;>{Leveque-ODEs}. The choice of $2$ for the exponent follows the
technical arguments given in \ref{app:A} that the operator $(1 -
L^2\nabla_H^2)^M$ has an associated Green's function with a characteristic
length-scale $L$ for the choice of $M\geq2$, such that $L$ may be interpreted as
a filter length-scale (e.g., the power spectrum falling off beyond some the
length-scale $L$). Fig.~\ref{fig:pre_mortem_L100_R025} shows the filter in
action. Note in particular that the zonal power spectrum of the large-scale
field at fixed latitude (averaged over all latitudes) shown in
Fig.~\ref{fig:pre_mortem_L100_R025}($d$) shows a power drop off to below
$10^{-7}\ {}^\circ\mathrm{C}^2$ that coincides with the choice of $L$ as long as
it is bigger than the Nyquist wavelength of the model (which is $2\times25 =
50\mathrm{km}$ for the dataset concerned).

The present implementation for the filter employs a Richardson iteration
\cite<e.g.,>{Richardson10, Trottenberg-et-al-Multigrid} with a pre-conditioning
with parameter $\gamma$, which was easier to implement in NEMO and suffices for
the cases considered; see \ref{app:A} for details. Within NEMO itself, a new
module called \verb|ldftra_split| was introduced to perform the filtering
procedure, leveraging a substantial amount of existing code for computing
diffusive trends associated with the discretized Laplacian operator (e.g.,
\verb|traldf_lap_blp|). If we are letting the relevant eddy-induced velocity
$\boldsymbol{u}^*_L$ act on the total tracer field (i.e. solving for
Eq.~\ref{eq:traadv}) then no further modifications except for the computation of
the relevant quantities are required. The addition of operations in NEMO is
roughly:
\begin{itemize}
  \item if splitting, call \verb|tra_ldf_split| provided by new module
  \verb|ldftra_split|, resulting in a large-scale thermodynamic field
  \verb|tsb_l|;
  \item use \verb|tsb_l| to recompute the related stratification variables
  (\verb|rab_b|, \verb|rn2b|, \verb|rhd|) \emph{after} the vertical physics step
  (which uses the full thermodynamic field \verb|tsb|);
  \item compute the slopes \verb|wslp[ij]| using the recomputed \verb|rhd| and \verb|rn2b|;
  \item \verb|rhd| is recomputed using full \verb|tsb| for ocean physics (NEMO
  already does this call by default);
  \item call GEOMETRIC routines in \verb|ldfeke|, where new \verb|wslp[ij]| are
  used to obtain eddy induced velocities \verb|aei[uv]|, and these are exposed
  to the tracer advection modules \verb|traadv| as usual.
\end{itemize}

The relevant data redirection are all performed in the NEMO time-stepping driver
module \verb|step|; see the source files provided as part of the open data
repository (search for \verb|tra_ldf_split| and \verb|tsb_l| in
\verb|step.F90|). 


\subsection{Model setup}

The above procedures were implemented in NEMO 4.0.5 (r14538), and tested in an
idealized zonally re-entrant channel model of the Southern Ocean. The model set
up is based on that of \citeA{Munday-et-al15}, with no continental barriers but
with a submerged ridge, and is a longer version of the channel model in
\citeA{Mak-et-al18}. The model domain is 9000 km by 2400 km by 3000 m in the
zonal, meridional and vertical direction, the choice of domain extent is such
that models with horizontal grid spacing of $100$, $50$, $25$ and $10$ km are
supported (denoted here by R100, R050, R025, R010 respectively). The model
employs a linear bottom drag with coefficient $r=1.1 \times 10^{-3}\ \mathrm{m}\
\mathrm{s}^{-1}$. Specifications of the submerged ridge (the domain is at 1500 m
depth at its most shallow), the purely zonal and time-independent wind stress
forcing (with peak wind stress at $\tau_0 = 0.2\ \mathrm{N}\ \mathrm{m}^{-2}$)
as well as the surface temperature restoring profile are the same as
\citeA{Munday-et-al15} and \citeA{Mak-et-al18}, adapted to the present domain.

The model takes a linear equation of state with temperature $\Theta$ as the only
thermodynamic variable. There is a constant vertical diffusivity $\kappa_v =
10^{-5}\ \mathrm{m}^2\ \mathrm{s}^{-1}$ everywhere in the domain, except in a
$300$ km sponge region towards the northern boundary where the value of
$\kappa_v$ is increased towards $\kappa_v = 10^{-3}\ \mathrm{m}^2\
\mathrm{s}^{-1}$ in the same manner described in \citeA{Munday-et-al15} and
\citeA{Mak-et-al18}. The choice of enhanced diffusivity differs to restoring
northern boundary conditions considered for example in
\citeA{Abernathey-et-al13} and \citeA{Youngs-et-al19}, and has an impact on the
residual meridional overturning circulation. Given the choice of a single
thermodynamic variable, no isoneutral diffusion \cite{Redi82, Griffies-et-al98}
is employed. All models detailed below employ a lateral hyperdiffusion in the
temperature field, chosen as $\kappa_4 = (1/12)U_e L_e^3$, where $U_e = 0.02\
\mathrm{m}\ \mathrm{s}^{-1}$ for all calculations, and $L_e = \Delta x = \Delta
y$ except for R100 where $L_e = 200\ \mathrm{km}$ to damp out the grid-scale
fluctuations. The hyper viscosities where employed follow the same prescription
but with $U_e = 0.1\ \mathrm{m}\ \mathrm{s}^{-1}$; the Laplacian viscosity for
R100 was chosen empirically. A summary of the key model parameters are given in
Table~\ref{tbn:param}.

Three sets of experiments were performed. One set has no GM-based
parameterization switched on (denoted $\kappa_{\rm gm} = 0$, neglecting units).
One set has the GEOMETRIC parameterization switched on and applied as standard
(employing Eq. \ref{eq:geom} and \ref{eq:ene-eq}). The final set and the
principal focus of the present work is the one with the GEOMETRIC
parameterization but with a splitting approach (Eq. \ref{eq:GEOMloc-e} and
\ref{eq:GEOM-gm}). All calculations were spun up from rest at the relevant
horizontal resolutions with $\kappa_{\rm gm} = 0$ to the end of model year 500,
and perturbation experiments were performed from start of model year 501 to end
of model year 810, where the time-averaging period is the last 10 years (start
of model year 801 to end of model year 810). All the GEOMETRIC-based
calculations here employ the same parameter values, given in
Table~\ref{tbn:param}; the choice of $\alpha$ and $\lambda$ is on the larger
side to previous reported works, but the value of $\lambda$ is not inconsistent
with the estimates for the Southern Ocean derived from an inverse calculation
\cite{Mak-et-al22b}.

\begin{table}
  \caption{Parameter values employed in the set of simulations.}
  \label{tbn:param}
  \centering
  \begin{tabular}{cc cccc c}
  \hline
  Parameter  & & R100 & R050 & R025 & R010 & Units\\
  \hline
   $\Delta x = \Delta y$ & grid spacing            & 100  &  50  & 25  & 10  & $\mathrm{km}$\\
   $\Delta t$            & time step size          &  60  &  40  & 20  & 10  & $\mathrm{mins}$\\
   $\nu_2$               & viscosity               & 5000 &  --- & --- & --- & $\mathrm{m}^2\ \mathrm{s}^{-1}$\\
   $\nu_4$               & hyper-viscosity         & ---  &  $5.2\times 10^{12}$ & $6.5\times 10^{11}$ & $6.2\times 10^{10}$ & $\mathrm{m}^4\ \mathrm{s}^{-1}$\\
   $\kappa_4$            & hyper-diffusivity       & $1.3\times 10^{13}$ & $2.1\times 10^{11}$ & $2.6\times 10^{10}$ & $1.6\times 10^9$ & $\mathrm{m}^4\ \mathrm{s}^{-1}$\\
   $L$                   & filter length scale     & --- &  100  & 100 & 100 & $\mathrm{km}$\\
   $\gamma$              & pre-conditioning param. & --- &   20  &  75 & 500 & ---\\
  \hline
  \end{tabular}
  \begin{tabular}{cc c c}
  \hline
  Parameter & & Common values & Units\\
  \hline
   $\alpha$         & eddy efficiency        & 0.06  & ---\\
   $\eta_E$         & energy diffusivity     & 500   & $\mathrm{m}^2\ \mathrm{s}^{-1}$\\
   $\hat{E}_0$      & minimum energy level   & 4.0   & $\mathrm{m}^3\ \mathrm{s}^{-2}$\\
   $\lambda^{-1}$   & dissipation time-scale & 80    & days\\
  \hline
  \multicolumn{3}{l}{}
  \end{tabular}
\end{table}


The choice of the GEOMETRIC parameters (principally $\alpha$ and $\lambda$,
respectively the coefficient related to the growth and dissipation of
parameterized eddy energy) were chosen as follows. We first perform a model
truth calculation R010 with $\kappa_{\rm gm} = 0$. We diagnose the total
explicit eddy energy as the sum of domain-averaged eddy kinetic energy
\begin{linenomath*}
\begin{equation}\label{eq:eke}
  \langle\mbox{EKE}\rangle = \frac{1}{V}\int_0^{L_x} \int_0^{L_y} \int_{-H(x,y)}^0 \frac{1}{2}\left(\overline{\boldsymbol{u}\cdot\boldsymbol{u}} - \overline{\boldsymbol{u}}\cdot\overline{\boldsymbol{u}}\right)\; \mathrm{d}z\ \mathrm{d}y\ \mathrm{d}x,
\end{equation}
\end{linenomath*}
where $\overline{(\cdot)}$ is a time average over the 10 year window, $V$ is the
volume of the computation domain, $-H(x,y)$ denotes the bathymetry, and
$L_{x,y}$ denotes the domain extents in the zonal and meridional directions. The
domain-averaged eddy potential energy is computed in density co-ordinates as
\begin{linenomath*}
\begin{equation}\label{eq:epe}
  \langle\mbox{EPE}\rangle = \frac{1}{V}\int_0^{L_x} \int_0^{L_y} \int_{\rho_b}^{\rho_t} \frac{g}{2\rho_0} \left(\overline{z^2} - \overline{z}^2\right)\; \mathrm{d}\rho\ \mathrm{d}y\ \mathrm{d}x,
\end{equation}
\end{linenomath*}
where $\rho$ is the density computed from the linear equation of state. Both of
the above quantities were computed from five-day averaged data output every five
days over the 10 year window. The $\alpha$ and $\lambda$ parameters were tuned
roughly so that the coarse resolution model R100 possesses similar total
(explicit and parameterized) eddy energy levels and total circumpolar transport
to the model truth calculation. The circumpolar transport is calculated as
\begin{linenomath*}
\begin{equation}\label{eq:transport}
  T_{\rm tot} = \frac{1}{L_x}\int_0^{L_x}\left(\int_0^{L_y}\int_{-H(x,y)}^0 \overline{u}\; \mathrm{d}z\; \mathrm{d}y\right)\; \mathrm{d}x,
\end{equation}
\end{linenomath*}
where $\overline{u}$ denotes the time-averaged zonal velocity.

For the results presented here all splitting was performed with the
aforementioned filter with fixed length-scale of $L=100\ \mathrm{km}$, and the
pre-conditioning parameter $\gamma$ was determined empirically as the value that
leads to a stable convergence of the Richardson iteration procedure in a Python
implementation of the filtering algorithm (see file provided in data
repository). For computation cost reasons, however, the filtering procedure is
only performed every model day, although the small-scale field (being the
residual of the total and large-scale field) is updated at every time-step. A
mild defense of this choice is that the large-scale spatial varying field is
expected to be slowly evolving in time \cite<cf.>{Rai-et-al21}, and an update at
every model time-step may not be strictly necessary. Sample experiments varying
the update period from every time-step to every model month did not lead to any
noticeable differences in the resulting conclusions of this article. With the
current choice of updating every model day, the extra run time given the same
amount of computational resources was empirically determined to be no more than
5\% for the R025 calculations.


\section{Results}\label{sec:results}

As described in previous works \cite<e.g.,>{Munday-et-al15, Mak-et-al18}, the
model is characterized by a flow arising from blocked $f/H$ contours because of
the submerged ridge, where $f$ is Coriolis parameter. The mean flow deflects
towards the north (or equator), and downstream of the ridge a standing meander
and substantial mesoscale eddy activity results; see for example the snapshots
of the surface relative vorticity shown previously in
Fig.~\ref{fig:xi_snap}($a,d$). The resulting Rossby deformation radius as
measured by $L_d = (2f)^{-1}\int N\; \mathrm{d}z$ \cite<cf.>{NurserBacon14}
varies in latitude from 5 to about 100 km from south to north, and has a
domain-average of around $50\ \mathrm{km}$; in that sense R010 is resolving
eddies except far into the south, R025 is eddy-permitting, R050 is barely
eddy-permitting, and R100 is non-eddy resolving. In the model truth R010 with
$\kappa_{\rm gm}=0$, the diagnosed total circumpolar transport
(Eq.~\ref{eq:transport}) is 115.0 Sv (where 1 Sv = $10^6\ \mathrm{m}^3\
\mathrm{s}^{-1}$), with a domain averaged specific total eddy energy of $0.0273\
\mathrm{m}^2\ \mathrm{s}^{-1}$ all in the explicit component (where about 20\%
of this is in the form of EKE). The R100 coarse resolution calculation employing
GEOMETRIC has the $\alpha$ and $\lambda$ roughly tuned to the aforementioned
transport and eddy energy values, and the resulting calculation with the
parameters given in Table \ref{tbn:param} has total circumpolar transport of
118.8 Sv and a domain averaged specified eddy energy of $0.0326\ \mathrm{m}^2\
\mathrm{s}^{-1}$, where the latter is a little high compared to the model truth.
About 32\% of the coarse resolution total eddy energy is in the explicit EPE,
and the remaining essentially in the parameterized eddy energy, with a
negligible amount in the explicit EKE.


\subsection{Scale-awareness in the energetics}

\begin{figure}
  \includegraphics[width=\textwidth]{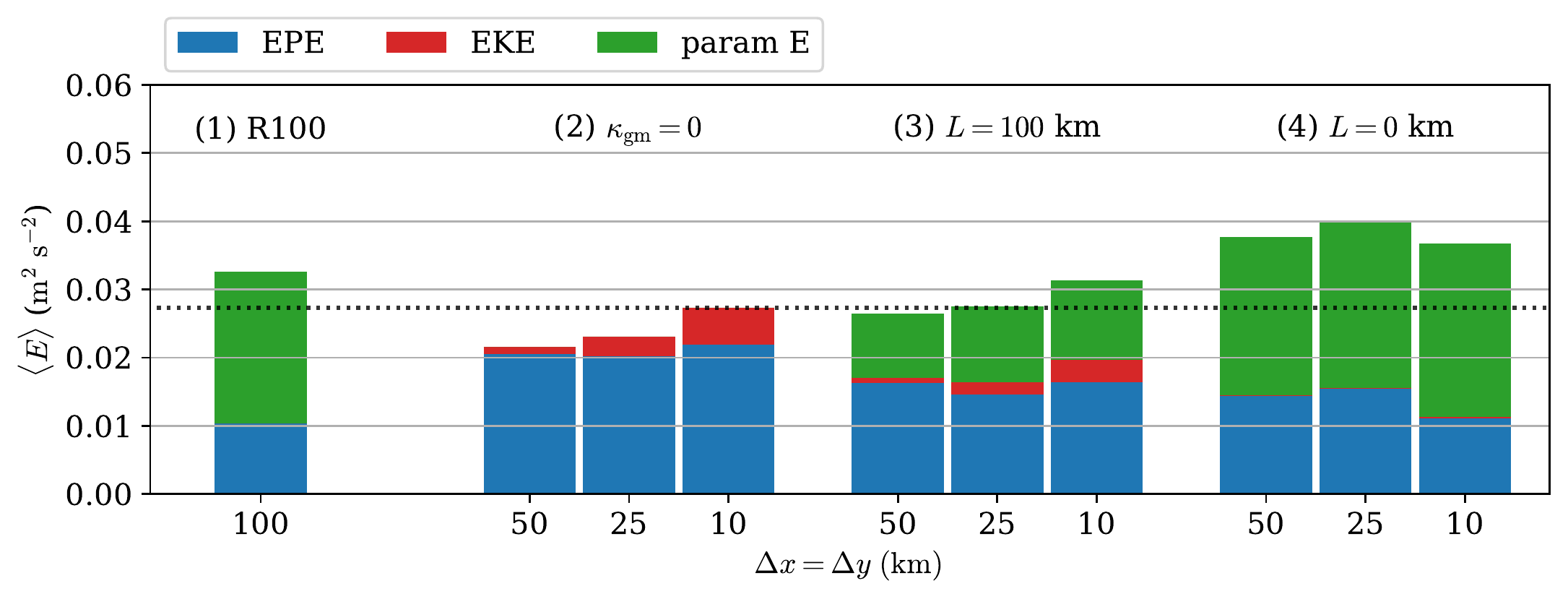}
  \caption{Domain-averaged total (explicit and parameterized) eddy energy levels
  and its decomposition for various calculations at varying resolution, grouped
  as (1) the coarse resolution calculation R100 at $\alpha=0.06$, $\lambda^{-1}
  = 80\ \mathrm{days}$, with parameters tuned so that the resulting eddy energy
  and total circumpolar transport are roughly the same as the model truth, (2)
  calculation with $\kappa_{\rm gm} = 0$, (3) calculations with GEOMETRIC at the
  same parameters as R100 and splitting, with filter scale $L=100\ \mathrm{km}$,
  (4) calculations with GEOMETRIC at the same parameters as R100 without
  splitting. The black dotted line denotes the total eddy energy level of the
  R010 $\kappa_{\rm gm}=0$ model truth calculation.}
  \label{fig:ene_decomp_100km}
\end{figure}

The information regarding the eddy energy across the set of calculations across
resolutions is summarized in Fig.~\ref{fig:ene_decomp_100km}. In the
$\kappa_{\rm gm} = 0$ calculations, the total eddy energy increases with
resolution as expected, with most of the explicit eddy energy as EPE that
remains roughly constant with increasing resolution, and a notable increase in
the explicit EKE. It should be noted that the R050 and R025 (the models in the
`eddy permitting' regime) with $\kappa_{\rm gm} = 0$ have a stratification that
is extended too deep compared to the model truth (e.g.,
Fig.~\ref{fig:hydrographics}$c$ later), with too large a transport compared to
the model truth.

In the calculations employing a filtering with the use of GEOMETRIC and
$\boldsymbol{u}^*_L$, there is a suggestion that the total eddy energy
(parameterized and explicit) is somewhat constant, indicating scale-awareness in
the eddy energetics, although a non-negligible amount of the total eddy energy
is in the parameterized component. The total explicit eddy energy component (in
both EKE and EPE form) is certainly lower in comparison to the corresponding
$\kappa_{\rm gm} = 0$ calculations, but is noticeably higher than the
calculations where GEOMETRIC is used as is with no filtering of fields
(particularly noticeable in the explicit EKE levels).

We note that there is a significant component of total eddy energy in the
parameterized component across the resolutions. A major contribution to the
parameterized eddy energy seems to be the presence of the standing meander. The
standing meander is persistent (albeit substantially weakened) under the
filtering, contributing to significant generation of parameterized eddy energy
that leads to damping of the explicit component via the associated eddy-induced
velocity, so that the parameterization ends up compensating for the reduced eddy
activity. Analogous model calculations without a ridge and so no standing
meander (not shown) also results in rough constancy of total eddy energy, but
with a substantially decreased parameterized component as resolution is
increased (cf. Fig.~\ref{fig:ene_decomp_schematic}$b$). The results suggest we
may want to consider filtering procedures that remove or reduce the projection
of the standing eddy onto the large-scale field; choices of the filter and the
definition of a large-scale field are further discussed in the conclusion
section.

We should note that the eddy energy dependence behavior discussed here seem to
be robust with reasonable variations of GEOMETRIC parameters $\alpha$ and
$\lambda$ with the present choice of filtering length scale $L = 100\
\mathrm{km}$, for sample sets of calculations that have been performed (not
shown), as long as the total circumpolar transport is roughly around 115 Sv
($\pm 10\%$ say).


\subsection{Mean state sensitivities}

To further investigate the impact of the splitting procedure in the set of
calculations, we take the R025 calculations as a working example to demonstrate
a few desirable features conferred to the model that results from the present
procedure; similar conclusions appear to hold in the R050 calculations (not
shown).

\begin{figure}
  \includegraphics[width=\textwidth]{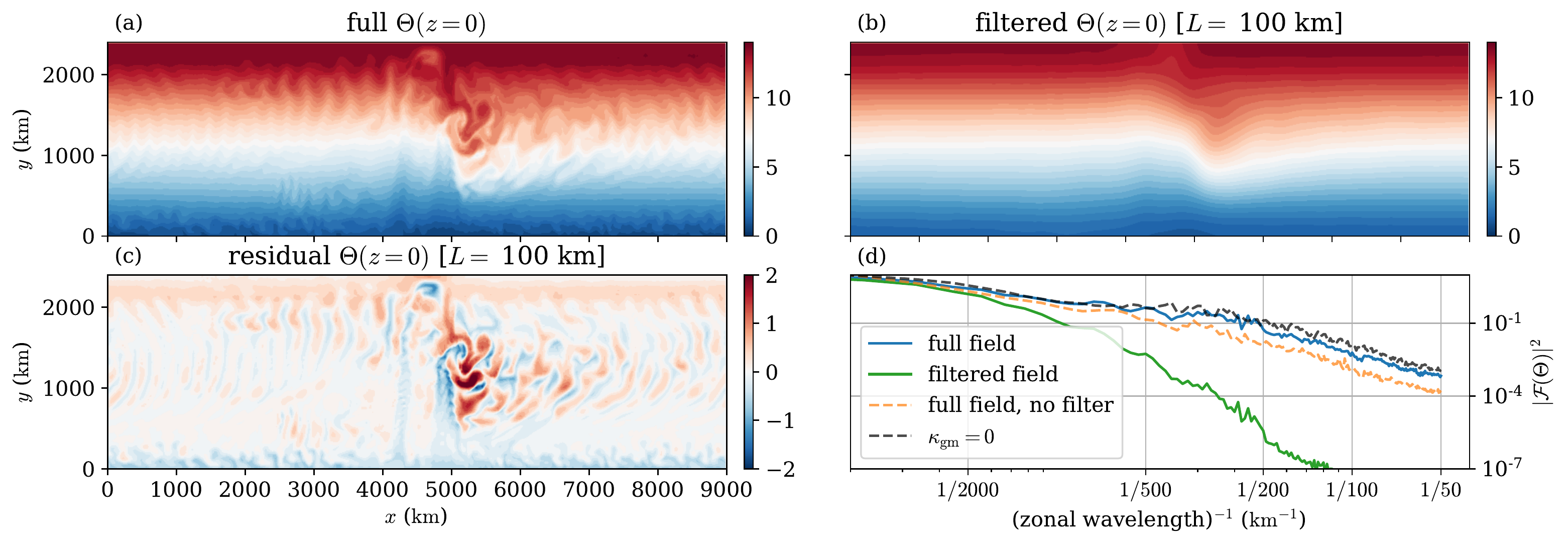}
  \caption{Snapshot of sea surface temperature R025 ($\Delta x = \Delta y = 25$
  km) calculation with GEOMETRIC ($\alpha=0.06$, $\lambda^{-1} = 80\
  \mathrm{days}^{-1}$) with splitting ($L=100\ \mathrm{km}$). ($a$) The
  diagnosed field, at the end of model year 810. ($b$) The diagnosed large-scale
  field with filter length $L=100\ \mathrm{km}$. ($c$) The small-scale field as
  a residual between the diagnosed original and the filtered field (so panel $a$
  minus $b$). ($d$) Zonal power spectra at fixed latitude averaged over all
  latitudes (in units of ${}^\circ\mathrm{C}^2$) for the diagnosed full field
  (panel $a$, blue), filtered field (panel $b$, green), and for comparison
  purposes, the analogous power spectrum for the full field for a calculation
  without filtering ($L=0\ \mathrm{km}$, orange dashed) and the $\kappa_{\rm
  gm}=0$ calculation (black dashed).}
  \label{fig:post_mortem_L100_R025}
\end{figure}

The reduced damping of the explicit eddies is apparent already in the snapshots
of the surface relative vorticity shown previously in
Fig.~\ref{fig:xi_snap}($d$). In the filtered variable,
Fig.~\ref{fig:post_mortem_L100_R025} shows a snapshot of the sea surface
temperature field diagnosed from a prognostic calculation of R025 with GEOMETRIC
and splitting active (in contrast to Fig.~\ref{fig:pre_mortem_L100_R025}, which
demonstrates the filtering procedure on a snapshot of a diagnosed tracer field
from a $\kappa_{\rm gm} = 0$ calculation). The model large-scale field has a
persistent zonal temperature gradient near the location of the ridge, arising
from the projection of the standing meander onto the large-scale field. A
substantial amount of activity is in the small-scales, as seen in
Fig.~\ref{fig:post_mortem_L100_R025}($c$), demonstrating that the filtering and
splitting procedure functions even in a prognostic setting (which is not
immediately obvious given the nonlinear feedbacks present in contrast to the
filtering applied diagnostically). To quantify the explicit activity, the zonal
power spectrum of the sea surface temperature at fixed latitudes averaged over
latitudes is shown in Fig.~\ref{fig:post_mortem_L100_R025}($d$). The power
spectrum density in the large-scale decreases (the green line) to a magnitude of
$10^{-7}\ {}^\circ\mathrm{C}^2$ at the filtering length-scale $L=100\
\mathrm{km}$ (cf. Fig.~\ref{fig:pre_mortem_L100_R025}). The power spectrum
density of the full field from the present calculation (the blue line) is
smaller across scales but otherwise reasonable compared to a calculation of R025
with $\kappa_{\rm gm}=0$ (the black dashed line), indicating a mild decrease in
explicit activity in R025 calculations with GEOMETRIC and the splitting
procedure. The analogous power spectrum for a calculation with no filtering (the
orange dashed line) is smaller than the two aforementioned cases by about an
order of magnitude for length-scales smaller than around $500\ \mathrm{km}$.

\begin{figure}
  \includegraphics[width=\textwidth]{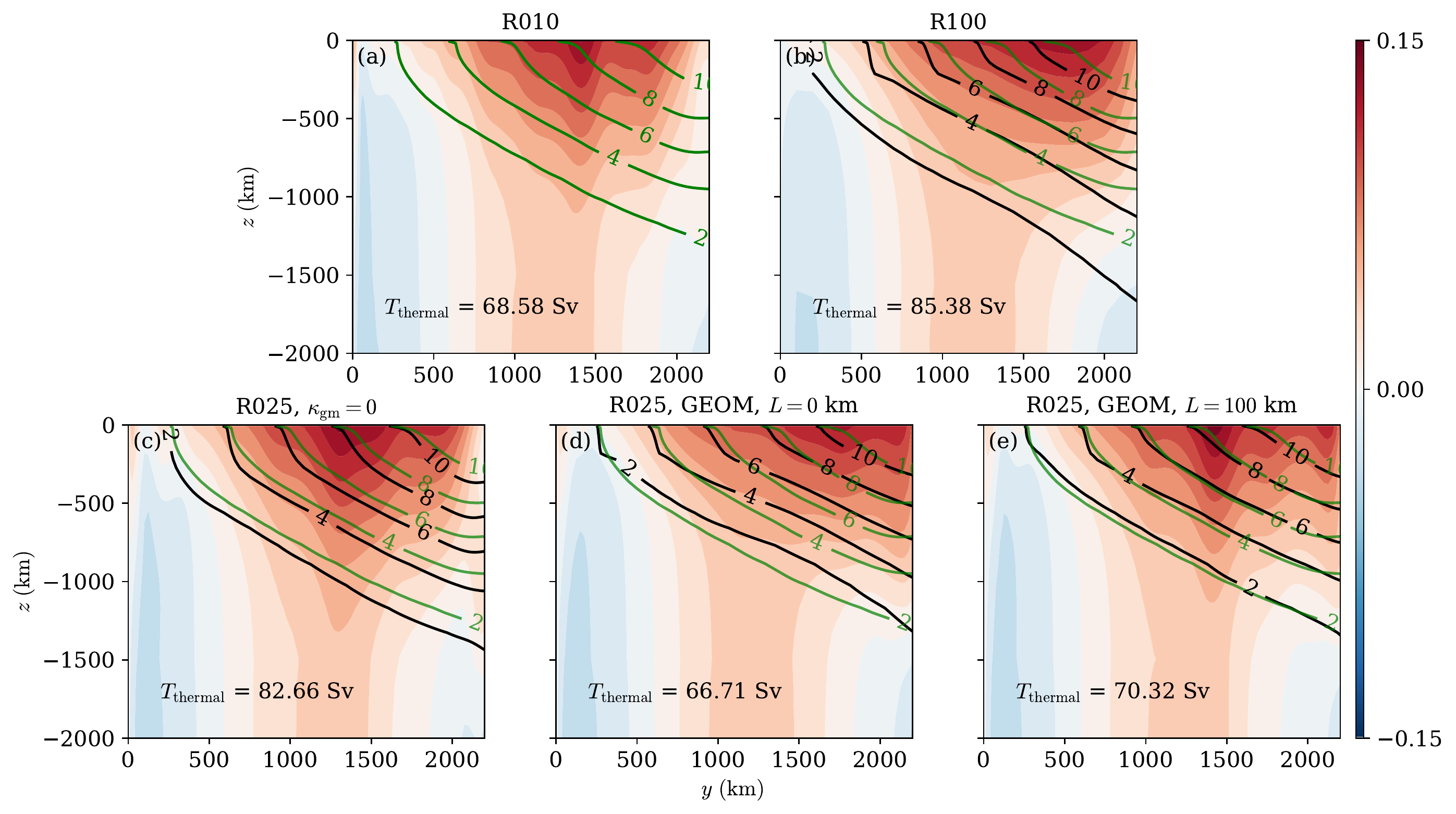}
  \caption{Zonal mean states for various calculations (with the northern
  boundary region with enhanced vertical diffusivity removed), showing the zonal
  averaged velocity (shading, in units of $\mathrm{m}\ \mathrm{s}^{-1}$) and
  isotherms (green contour lines for model truth R010, black contour lines
  otherwise, in units of ${}^\circ\ \mathrm{C}$). ($a$) The model truth R010
  with $\kappa_{\rm gm}=0$. ($b$) The coarse resolution calculation R100 at
  $\alpha=0.06$, $\lambda^{-1} = 80\ \mathrm{days}$, with parameters tuned so
  that the resulting eddy energy and total circumpolar transport are roughly the
  same as the model truth. ($c$) R025 with $\kappa_{\rm gm} = 0$. ($d$) R025
  with GEOMETRIC at the same parameters as R100, using no filtering and using
  the full-scale $\boldsymbol{u}^*$. ($e$) R025 with GEOMETRIC at the same
  parameters as R100, with filter at $L=100\ \mathrm{km}$, and using the
  large-scale $\boldsymbol{u}^*_L$. The corresponding $T_{\rm thermal}$ from
  Eq.~(\ref{eq:therm}) is marked onto the figure as a quantitative measure
  relating to the stratification.}
  \label{fig:hydrographics}
\end{figure}

A sample set of the zonal mean temperature and zonal velocity profiles is given
in Fig.~\ref{fig:hydrographics}. It can be seen here that while R100 has a
comparable total zonal transport (of around 115 Sv) to the model truth R010 with
$\kappa_{\rm gm} = 0$, their stratification differs, with the R100 calculations
having a stratification extending deeper (Fig.~\ref{fig:hydrographics}$a,b$).
This observation is reflected in the diagnosis of a thermal wind or baroclinic
transport
\begin{linenomath*}
\begin{equation}\label{eq:therm}
  T_{\rm thermal} = T_{\rm tot} - T_{\rm bot}, \qquad T_{\rm bot} = \frac{1}{L_x}\int_0^{L_x}\left(\int_0^{L_y}\int_{-H(x,y)}^0 \overline{u}(z=-H)\; \mathrm{d}z\; \mathrm{d}y\right)\; \mathrm{d}x,
\end{equation}
\end{linenomath*}
where $T_{\rm tot}$ is given by Eq.~(\ref{eq:transport}), and the second term is
the transport associated with the bottom flow (in practice $\overline{u}(z=-H)$
being taken at the first wet point above the modeled bathymetry). R100 with the
present choice of GEOMETRIC parameters has $T_{\rm thermal} = 85.4$ Sv, while
the model truth R010 has $T_{\rm thermal} = 68.9$ Sv. The stronger
depth-independent component with the increased resolution as mesoscale eddies
are resolved is consistent with the findings of \citeA{Yankovsky-et-al22}.

For the R025 calculations, we note that the $\kappa_{\rm gm} = 0$ calculations
(Fig.~\ref{fig:hydrographics}$c$), while showing the most variability
(Fig.~\ref{fig:xi_snap}$a$ and Fig.~\ref{fig:post_mortem_L100_R025}$d$), has a
deeper zonal mean stratification profile compared to the model truth R010,
consistent with $T_{\rm thermal} = 82.7$ Sv in this calculation. The calculation
employing GEOMETRIC but no splitting (Fig.~\ref{fig:hydrographics}$d$) leads to
a zonal mean stratification that has mismatches in the deeper regions, although
the thermal wind transport is comparable to the model truth, with $T_{\rm
thermal} = 66.7$ Sv. Note the substantially weaker variations in the zonal mean
zonal velocity in the GEOMETRIC without splitting calculation, consistent with
the strong damping of the present model seen in the snapshots (cf.
Fig.~\ref{fig:xi_snap}$c$). The calculation with GEOMETRIC and splitting
(Fig.~\ref{fig:hydrographics}$e$) leads to a zonal mean stratification, a
baroclinic transport of $70.3$ Sv and a zonal mean flow that is meridionally
confined, all of which agrees well with the model truth calculation. Further,
the zonal mean zonal velocity also possesses more spatial fluctuations,
consistent with the reduced damping of the explicit eddies. The splitting
approach in this calculation allows the extra flattening of isopycnals arising
from the GM-based GEOMETRIC scheme, but with only very mild damping of the
explicit fluctuations. The similar diagnosed values of $T_{\rm thermal}$ in the
R025 cases considered is likely due to the better resolved standing eddy over
the ridge; similar observations are seen in the analogous R010 calculations,
though less so in the R050 set of calculations.

The calculation with splitting appears to allow both the explicit and
parameterized eddy-mean feedbacks to be present, and provides the most
satisfactory representation of the zonal mean stratification and zonal mean
flow, while maintaining a degree of variability not present with standard
implementations of GM-based schemes. It should additionally be noted that the
agreement in the stratification appears to \emph{require} the splitting
procedure to be active. In our sample calculations of R025 employing GEOMETRIC
(not shown), we find that for fixed $\lambda$ (since changes in the mean state
from $\lambda$ can be offset accordingly by inverse changes in $\alpha$), we can
certainly tune $\alpha$ so that the total or baroclinic transport agree with the
model truth, but biases in the stratification appear to be persistent (e.g.,
inspecting the isopyncal contours visually). In that sense, there is some
evidence lending support to the conclusion that the splitting approach really
does lead to a better represented mean state -- a mean state that standard
procedures struggle to replicate by tuning. Analogous calculations in a model
without a ridge results in a similar conclusion, but with the R025 $\kappa_{\rm
gm}=0$ calculation having an even deeper stratification (not shown), attributed
to the fact there is no longer form stress contributions from the standing eddy
to the momentum budget.

\begin{figure}
  \includegraphics[width=\textwidth]{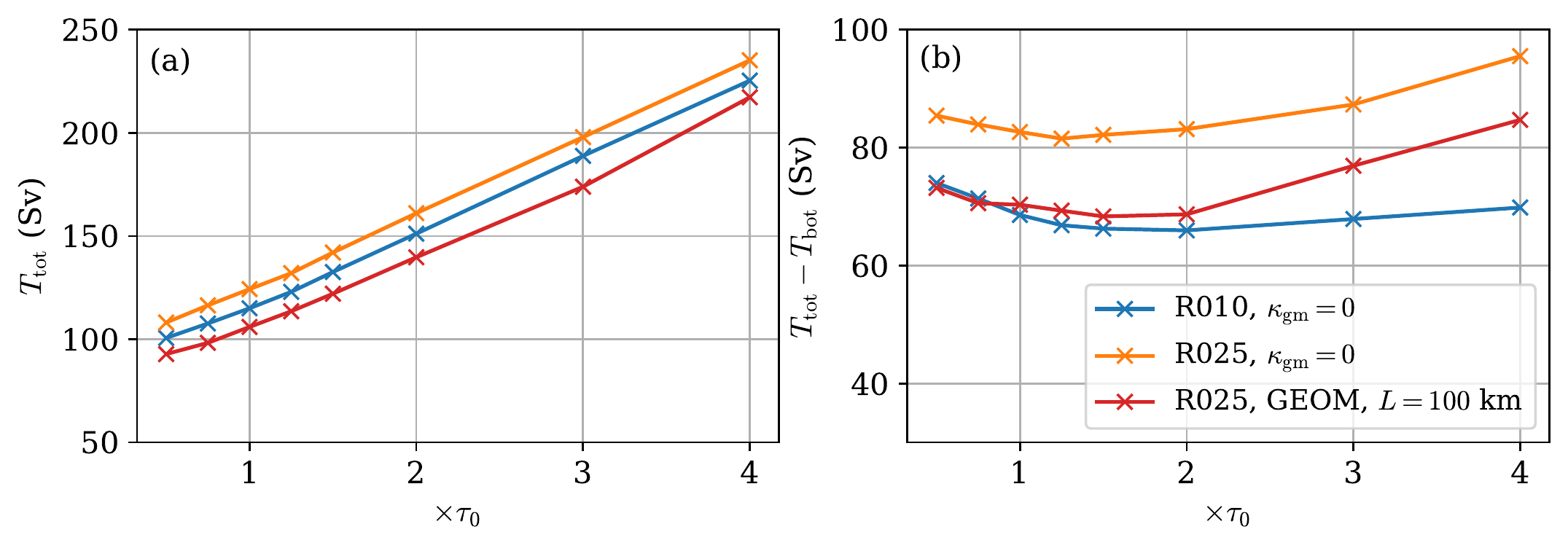}
  \caption{Circumpolar transport diagnostics at varying wind stress forcing
  ($1\times\tau_0$ being the calculation at control wind stress), for model
  truth R010 with $\kappa_{\rm gm}=0$ (blue), R025 with $\kappa_{\rm gm}=0$
  (orange), and R025 with GEOMETRIC and filtering active ($L=100\ \mathrm{km}$,
  red). ($a$) Total circulation transport $T_{\rm tot}$
  (Eq.~(\ref{eq:transport}). ($b$) Baroclinic transport $T_{\rm tot} - T_{\rm
  bot}$ (Eq.~\ref{eq:therm}).}
  \label{fig:transport}
\end{figure}

A set of wind perturbation experiments were performed by multiplying the zonally
symmetric wind stress by a constant factor, again starting from the start of
model year 501 to the end of model year 810, for the model truth R010 with
$\kappa_{\rm gm} = 0$, R025 with $\kappa_{\rm gm} = 0$, and R025 with GEOMETRIC
using the splitting procedure. Fig.~\ref{fig:transport} shows the diagnosed
total transport and thermal wind transport calculated from the time-averaged
data in the same analysis period (start of model year 801 to end of model year
810). The model truth R010 with $\kappa_{\rm gm} = 0$ has a mildly decreasing
baroclinic transport with increasing wind before rebounding somewhat at very
large wind forcing (four times the control wind stress). The sensitivity of
transport was previously observed in the shorter channel model of
\citeA{Mak-et-al18}, and possibly arises from the use of an enhanced vertical
diffusivity in the northern boundary, leading to a residual overturning in the
opposite direction to the usual one \cite{Youngs-et-al19}.

The R025 case with $\kappa_{\rm gm} = 0$ follows the decreasing trend in the
baroclinic transport of the model truth for moderate winds, but increases again
at larger winds. The overall sensitivity in the thermal wind transport with
changes in wind stress however is substantially weaker than the calculations in
a non-eddying calculation R100 using a $\kappa_{\rm gm} =$ constant \cite<not
shown, but cf. Fig. 1 of>{Mak-et-al18}, likely because of the improved
representation of form stress when some eddies are permitted
\cite<cf.>{Munday-et-al13}, and also because the standing meander is better
resolved \cite{Stewart-et-al22}. For the R025 case employing GEOMETRIC and
splitting, the thermal wind transport follows the trend of the model truth well
up to twice the control wind forcing, but increases significantly at larger wind
forcing. To rationalize the observed behavior, note that both the transient and
standing eddy play a role in the momentum balance via their role in vertical
momentum transport. With splitting, where there is a degree of separation
between the explicit and parameterized eddies, with the latter increasing with
wind forcing as describe by GEOMETRIC, the explicit standing eddy is still only
partially resolved, leading to differing sensitivity to that observed in the
model truth. On the other hand, in the cases without splitting \cite<e.g. R100;
cf.>{Mak-et-al18}, or a case considered by accident here with splitting but
where parameterized eddy energy advection by the mean flow was essentially
inactive, the parameterized component intrudes on the explicit eddy and takes
over its role to various degrees in the momentum budget, leading to different
sensitivities (almost eddy saturated, with very weak dependence of thermal wind
transport with increases in wind stress; not shown). The present observations
further highlight the importance and ongoing challenges in representing the
eddy-mean feedback from the standing eddy, be it explicitly or via a
parameterization.


\section{Conclusions and outlooks}\label{sec:conc}

An aim of this work is to re-examine whether it is possible to achieve a more
physical representation of mesoscale eddy feedback with existing
parameterization schemes, for use in the next generation of eddy-permitting
ocean models. To that end, the primary question we consider here is whether the
GM-based version of GEOMETRIC is scale-aware in the energetics, in that the
total (parameterized plus explicit) eddy energy remains roughly constant over
different spatial resolutions, without recalibration of parameters. Within the
context of an idealized re-entrant channel model as a representation of the
Southern Ocean, the diagnosed total eddy energy levels shown in
Fig.~\ref{fig:ene_decomp_100km} provides evidence that the GM-based version of
GEOMETRIC is in fact scale-aware, as long as a splitting approach based on
filtering is utilized so that the parameterized component does not completely
dominate the total eddy feedback. As the explicit eddies become stronger with
increased horizontal resolution, the resulting explicit feedback is reflected in
the large-scale state, which affects the growth of the parameterized eddy energy
(Eq.~\ref{eq:GEOMloc-e}), modifying the resulting $\kappa_{\rm gm}$
distributions such that the parameterized eddy component ``makes way'' for the
explicit eddy component, and vice-versa. While we might have expected that the
splitting approach would allow the parameterized and explicit eddy component to
exist side by side, it was still not obvious that the GM-based version of
GEOMETRIC would necessarily be scale-aware, so the present finding is by no
means trivial. The scale-awareness is demonstrated for one choice of the
GEOMETRIC parameters (see Table~\ref{tbn:param}), where the choice was chosen so
that the resulting total transport and total eddy energy levels roughly
coincides with the model truth calculation. The scale-awareness result seems to
be robust across sample sets of calculations (with different GEOMETRIC
parameters, and analogous model calculations in the absence of a topographic
ridge) as long as the resulting total transport of the coarse resolution model
was roughly that of the model truth, differing only in the total eddy energy
levels and in the exact partition of the explicit and parameterized eddy energy.
In addition to scale-awareness, the splitting approach leads to various
improvements to eddy permitting models that are also non-trivial, such as
reducing bias in the mean stratification profiles without sacrificing the
dynamical fluctuations (e.g., Fig.~\ref{fig:hydrographics}$e$), and improving on
the transport values and possibly on its sensitivities (e.g.,
Fig.~\ref{fig:transport}$b$).

The principle behind the splitting approach advocated here is summarized in
Fig.~\ref{fig:schematic}: we would like to split out a large-scale non-eddying
state from the full state via a filter, compute an eddy-induced velocity
$\boldsymbol{u}^* = \nabla\times \left(\boldsymbol{e}_z \times \kappa_{\rm gm}
\boldsymbol{s}\right)$ from the resulting large-scale non-eddying state, and
apply the resulting $\boldsymbol{u}^*$ only to the large-scale state. The
resulting $\boldsymbol{u}^*$ as computed is an inherently large-scale and
smaller magnitude object, which avoids damping the explicitly resolved
small-scale eddies, and is closer to the original intention and derivation of
\citeA{GentMcWilliams90} via a Reynolds averaging procedure. For this first
work, two simplifications were made in the proposed splitting approach that
differs from the schematic given in Fig.~\ref{fig:schematic}, namely (1) a
horizontal filtering rather than a filtering along-isopycnals, and (2) the
eddy-induced velocity $\boldsymbol{u}^*$ acts on the full tracer field. The
first is partly justified in that we are in the small aspect ratio regime where
the horizontal average serves as a reasonable first approximation to
along-isopycnal averaging; the horizontal average utilizes a filtering (see
Eq.~\ref{eq:diff_imp}) with a well-defined length-scale (fixed to be $L=100\
\textnormal{km}$ in this work). The second is for simplicity, and we expect that
since $\boldsymbol{u}^*$ is a large-scale field with suitable choices of the
filter, the small-scales are already somewhat passively advected by the
eddy-induced velocity, so would allow scale-awareness to emerge. The
scale-awareness in the eddy energetics as shown in
Fig.~\ref{fig:ene_decomp_100km} with the present approach we would expect to be
improved by further refinements to the procedure.

The splitting procedure, while still employing a variant of the more standard
isotropic advection provided by the GM scheme, removes a modeling need for a
resolution function \cite<e.g.,>{Hallberg13}, which modifies the value of
$\kappa_{\rm gm}$ directly, with impact on the resulting eddy-induced velocity
$\boldsymbol{u}^*$; for this work and our particular choice of filter, the
resolution scale is replaced by the definition of filtering length-scale $L$. We
also note that the existing splitting approach is agnostic to the choice of the
GM-based scheme itself. Here we happen to choose the GM-based GEOMETRIC scheme,
so there are some subtleties with the parameterized eddy energetics that one has
to be careful about (see \S\ref{sec:energetics}). The analysis however places no
restriction on the exact specification of $\kappa_{\rm gm}$, and could be used
with other GM-based variants \cite<e.g.,>{Visbeck-et-al97, Treguier-et-al97,
EdenGreatbatch08, Jansen-et-al15a, Jansen-et-al15b, Jansen-et-al19}. In
addition, the splitting approach is not mutually exclusive with backscatter
parameterizations \cite<e.g.,>{Zanna-et-al17, Bachman19, Jansen-et-al19,
Juricke-et-al20}, and some backscatter would even be preferable to energize the
explicit flow and capture physical backscatter mechanisms. We would however make
a note that, since the explicit eddies themselves are not strongly damped in the
splitting approach, the degree of backscatter likely does not need to be so
large. It would be of interest to see if the GM-based GEOMETRIC scheme (or
indeed other existing mesoscale eddy parameterizations) in the presence of
backscatter would still be scale-aware in the energetics, but this is beyond the
scope of the present work.

In the present work we made the choice of studying scale-awareness in relation
to the eddy energetics, when others have considered the total (mean and eddy)
energetics \cite<e.g.,>{Jansen-et-al19}. We have analyzed the mean kinetic
energy and see that similar conclusions regarding scale-awareness appear to
hold, although we note that in the present model the domain-integrated mean
kinetic energy is smaller than the total eddy energy by about an order of
magnitude. Similar observations in the ratio between kinetic energy of
depth-averaged component and the residual baroclinic component to
\citeA{Yankovsky-et-al22} also hold, in both the mean and eddy component (not
shown). However, compared to the mean kinetic energy, the computation for mean
potential energy is more troublesome, since the reference for available
potential energy is in general difficult to compute \cite<e.g.,>{Tailleux13,
HieronymusNycander15, SuIngersoll16}, particularly for data from a $z$-level
model. In this work we made the choice to focus on the eddy energetics because
the references are well-defined for eddy energetics, and present only results
for eddy energetics for consistency reasons.

We make some further comments to our two simplifications made in this present
work. Starting with the splitting procedure, the present work employs a filter
per horizontal level given by Eq.~\ref{eq:diff_imp}, with a well-defined
length-scale $L$, where for this work we took $L=100\ \textnormal{km}$. With a
decreased $L$ (e.g., $L=50\ \textnormal{km}$), the splitting becomes incomplete
and we somewhat return into the regime where the parameterized component
dominates. For larger $L$ (e.g., $L=200\ \textnormal{km}$), the parameterized
component becomes weak in the intermediate resolution calculations R025 and
R050, leading to a deviation from total eddy energy constancy. For the present
choice of filter, two elliptic solves are required, but since the filtering
procedure is only carried out every model day, the computation costs are rather
minimal and empirically determined to be no more than 5\% extra run time in the
R025 calculations, possibly improved further with a better solver. Sample
calculations show no noticeable impact on the resulting calculations even if the
splitting procedure was carried out once every month, which could be attributed
to the fact that the large-scale field is unlikely going to evolve on a fast
time-scale.

On a side point relating to large-scale field evolving on a slower time-scale,
the Reynolds averaging procedure in deriving the GM-scheme would be more
appropriate using a time-based averaging. Employing a spatial filter as we have
done here is implicitly assuming that there is some equivalency between a time
and space averaging, and is only really valid under rather specific assumptions.
There is indeed a mix of averages used interchangeably here, and while this may
not be an unreasonable choice to make from a practical point of view (and is
indeed used by other works to do with eddy parameterization in the literature),
it is ultimately a choice that should be further examined and refined, with its
impacts quantified if possible.

The quantitative details are almost certainly dependent on how the large-scale
field is defined. The present work utilizes a fixed choice of filtering
length-scale $L$, but it is perfectly possible to have this as a parameter that
varies in space and time, since the problem is intrinsically based on solving a
diffusion-like equation. For example, choices of $L$ that are some multiple of a
local Rossby deformation radius should be possible, although some modifications
would presumably be required so that the pre-conditioning parameter $\gamma$
required for numerical convergence varies with the choice of filtering
length-scale. Investigations with other types of spatial filters or
coarse-graining operators \cite<e.g.,>{Aluie19, Grooms-et-al21} or some sort of
dynamics based splitting are possible and should be considered, although we note
that there are computation considerations that should be taken into account
(e.g. halo sizes when parallel computation is involved, re-computation of the
filtering kernel if the filtering length is to vary in time). It should also be
advantageous to define some sort of operator that further removes the projection
of the standing eddy onto the large-scale field. Analogous numerical
calculations (not shown) removing the ridge from the modeled system, thereby
removing the presence of the standing eddy, still results in rough
scale-awareness in the total eddy energy as in Fig.~\ref{fig:ene_decomp_100km},
but with a more significant explicit component and a much weaker parameterized
component as resolution is increased (cf.
Fig.~\ref{fig:ene_decomp_schematic}$b$). Beyond a horizontal filtering as we
have considered, it might be more satisfactory to consider ways to define a
large-scale isopycnal surface in line with the schematic in
Fig.~\ref{fig:schematic}, via rolling averages in the co-ordinates or other
computation approaches \cite<e.g.,>{Kafiabad22, KafiabadVanneste23}, but is
beyond the scope of the present work.

Regarding the choice here that the eddy-induced velocity $\boldsymbol{u}^*$ is
applied to the full tracer field, this is largely for simplicity reasons, with
the assumption that the small-scale is advected by $\boldsymbol{u}^*$ somewhat
passively. To have the eddy-induced advection acting only on the large-scales,
however, requires an equation governing the large-scale field that remains to be
rigorously determined. While it is certainly possible to form a large-scale
equation from the usual Reynolds averaging procedure, the problem here is in
determining which velocity field should be used: should it be the total
velocity, the filtered velocity, the velocity associated with the filtered
thermodynamic field (via geostrophic balance for example), or something else,
with additions of the eddy-induced velocity as appropriate? The highlighted
issues remain to be settled.

The present work employs a linear equation of state with temperature as the only
thermodynamic variable, and the filtering procedure is applied directly on the
temperature field per horizontal level. Care needs to be taken when utilizing a
nonlinear equation of state, since the filtered density is not the same as the
density of the filtered thermodynamic variables. We propose that, with a
nonlinear equation of state, it should be the neutral density that is to be
filtered, from which we calculate the associated eddy-induced velocity
$\boldsymbol{u}^*$, and apply that to the tracer equations. This might be
reasonable if we continue with the approximation that $\boldsymbol{u}^*$ is to
be applied to the full tracer field, but care presumably needs to be taken for
points raised in the previous paragraph if $\boldsymbol{u}^*$ is to act on
to-be-determined large-scale tracer equations. One aspect that should be
examined in more detail is the degree of non-adiabaticity introduced by the
present procedure. For the present case where $\boldsymbol{u}^*$ is applied to
the full tracer equation we suspect the non-adiabaticity to be rather small. A
slightly more problematic aspect concerns isoneutral diffusion \cite{Redi82,
Griffies98}. In this model with a single thermodynamic variable there is no
isoneutral diffusion, and it is not clear with a nonlinear equation of state
whether the diffusion should be along the isoneutral directions of the full
isopycnals, or the large-scale isopycnals. The latter is likely going to
introduce some cross isopyncal transport, while the former simply requires
storing and/or re-computing the isopycnal slopes and is likely a `safer' default
option. Both of these issues should be addressed with a comprehensive assessment
quantifying the associated impacts, but is beyond the scope of the present work.

From a practical point of view, it would be of interest to further test out the
splitting procedure in different and/or more realistic ocean models, but of
course bearing in mind various aforementioned subtleties that we should check
for in the related investigations. An extension of the investigation in
\citeA{Ruan-et-al23} assessing the impacts of the splitting procedure to
physical and biogeochemical responses in an idealized gyre model is ongoing. For
more realistic global configuration models, it is known that there are various
issues with the physical response with eddy permitting ocean models particularly
in the Southern Ocean, such as the circumpolar transport being too large,
impacting on the other connected components in the ocean such as the Southern
Ocean gyres, sea ice, and on the resulting tracer transports
\cite<e.g.>{Hewitt-et-al20}. It would be of interest to see if such a splitting
procedure proposed here with the use of a GM-based GEOMETRIC scheme is able to
reduce the known biases, given the efficacy of the scheme demonstrated in the
present idealized Southern Ocean only ocean model, and is subject of some
planned future works.


\section*{Data Availability}

This work utilizes the Nucleus for European Modelling of the Ocean model
\cite<>[v4.0.5, r14538]{NEMO_man}. The instructions on setting up the numerical
model, implementation of algorithm into NEMO, model data, and scripts used for
generating the plots in this article are available through the repository at
\citeA{Mak23}.

\appendix

\section{Technical details relating to the filter}\label{app:A}

The filtering procedure employed in this work is associated with the equation
\begin{linenomath*}
\begin{equation}\label{eq:helmholtzM}
  (1 - L^2\nabla_H^2)^M\Theta_L^n = \Theta^n,
\end{equation}
\end{linenomath*}
where $\Theta^n$ is the total field to be filtered at time-step $n$, $\Theta_L$
is the filtered field, $L$ is the length-scale of choice, and in this work we
take $M=2$. If we have $M=1$ then the above is essentially a pseudo
time-stepping of the diffusion equation with a backward Euler scheme. The choice
of $M=2$ is made for several properties, such as an interpretation as a filter
where the radial spectral power density decreases after a specified
length-scale, or through a convolution with a kernel that has decreasing support
after specified length-scale \cite<e.g.,>{Whittle63, Lindgren-et-al11}, and is
closely related to the Mat\'ern auto-covariance.

The operator $(1 - L^2\nabla_H^2)$ is positive definite and symmetric, and the
resulting system could be readily treated with standard methods \cite<e.g.,
conjugate gradient;>{Leveque-ODEs}. In this work we consider employing
Richardson iteration \cite<e.g.,>{Richardson10, Trottenberg-et-al-Multigrid},
i.e.
\begin{linenomath*}
\begin{equation}\label{eq:app_eq1}
  \Theta^{k+1} = \Theta^n + L^2 \nabla^2_H \Theta^k, \qquad \Theta^{k=0} = \Theta^n,
\end{equation}
\end{linenomath*}
where $\Theta^n$ is the input field at time-step $n$ and is kept fixed as
iteration number $k$ is varied, as it is easier to implement in NEMO and is
sufficient for the cases considered. If the iteration by (\ref{eq:app_eq1})
converges to some $\Theta^*$, then we have the solution to
Eq.~(\ref{eq:helmholtzM}) for the $M=1$ case. We repeat the outlined iteration
procedure once more to obtain the solution $\Theta^{n+1}$ for the $M=2$ case.

The method above converges if the matrix 2-norm of the discretization of $(1 -
L^2\nabla_H^2)$ is sufficiently small, which is dependent on the choice of $L$
and the grid spacing $\Delta x = \Delta y$ through the discretization of
$\nabla_H^2$. To aid convergence, we consider a pre-conditioning of
Eq.~(\ref{eq:app_eq1}) given by \cite<e.g.,>{Trottenberg-et-al-Multigrid}
\begin{linenomath*}
\begin{equation}\label{eq:app_eq2}
  \Theta^{k+1} = \frac{1}{\gamma}\Theta^n + \left[\frac{\gamma - 1}{\gamma} + \frac{L^2}{\gamma} \nabla^2_H\right] \Theta^k, \qquad \Theta^{k=0} = \Theta^n,
\end{equation}
\end{linenomath*}
where some $\gamma \geq 1$ is an acceleration parameter chosen to ensure
convergence, motivated by $\gamma[I/\gamma - (L^2/\gamma) \nabla^2_H]$, where
the role of $\gamma$ is to reduce the norm of the operator. A larger $\gamma$ is
required for convergence, though it also tends to reduce the rate of
convergence. The actual values used are determined empirically, and in the
present calculations are given in Table~\ref{tbn:param}.

In this work, the convergence criteria for the Richardson iteration is that the
global supremum norm is below some tolerance, i.e.
\begin{linenomath*}
\begin{equation}\label{eq:app_eq3}
  \|\Theta^{k+1} - \Theta^k\|_{\infty} = \max_{(x,y)}|\Theta^{k+1}(x,y) - \Theta^k(x,y)| < \epsilon.
\end{equation}
\end{linenomath*}
For this work, since we are only filtering the temperature field, we take
$\epsilon = 0.001\ {}^\circ\mathrm{C}$, and all reported calculations converged
within 500 iterations. With the current choices, the extra computation costs are
rather minimal, empirically determined to be no more than 5\% extra run time,
and possibly improved further with a better solver.

\acknowledgments

This research was funded by both the RGC General Research Fund 16304021 and the
Center for Ocean Research in Hong Kong and Macau, a joint research center
between the Qingdao National Laboratory for Marine Science and Technology and
Hong Kong University of Science and Technology. We thank the referees Peter
Gent, Elizabeth Yankovsky and Julie Deshayes for their comments that helped
clarify some scientific/algorithmic points and improve the presentation of the
article.


%
%


%
%
%
%
%

\end{document}